# A STOCHASTIC INVESTMENT MODEL FOR ACTUARIAL USE

# IN SOUTH AFRICA

By Ş Şahin and S Levitan

## ABSTRACT

In this paper, we propose a stochastic investment model for actuarial use in South Africa by modelling price inflation rates, share dividends, long term and short-term interest rates for the period 1960-2018 and inflation-linked bonds for the period 2000-2018. Possible bi-directional relations between the economic series have been considered, the parameters and their confidence intervals have been estimated recursively to examine their stability and the model validation has been tested. The model is designed to provide long-term forecasts that should find application in long-term modelling for institutions such as pension funds and life insurance companies in South Africa

## CONTACT DETAILS

Şule Şahin, Department of Mathematical Sciences, Institute for Financial and Actuarial Mathematics, University of Liverpool, Liverpool, UK; E-mail: Sule.Sahin@liverpool.ac.uk   and Department of Actuarial Sciences, Hacettepe University, Ankara, Turkey

Shaun Levitan, Colourfield, Johannesburg, South Africa, E-mail: shaun@colourfield.co.za

# 1. INTRODUCTION

1.1 Economic scenario generators (ESG) are computer-based models to be used for joint simulation of integrated economic series such as price inflation, interest rates, gross domestic product, stock prices and foreign-exchange rates by including the interaction of the series. These stochastic models provide the necessary economic output for modelling of liabilities that financial institutions would require. It is expected that an ESG should include the production of simulation results which reflect certain financial variables and relevant view of the economy. It should include some extreme but plausible results and produce scenarios reflecting realistic market dynamics.

1.2 Campbell et al. (2016) present an extensive review on the nature of ESGs by providing definitions, how they evolved and how they address the needs in the insurance and pension industries. They also discuss the required features of a good ESG as well as providing a detailed guidance on crucial aspects of the financial market specification, model calibration and validation. The increased utilisation of ESGs are derived by two common applications: real world and market-consistent models. Real world models are concerned with future potential paths of economic variables and their potential influence on capital and solvency while the market-consistent models are more concerned with mathematical relationships within and among financial instruments.

1.3 The range of the applications of ESGs might require the use of different types of models which leads a discussion between the use of the real world and market consistent models. The real world simulation aims to capture market dynamics, risks and returns in a way that the financial institutions or insurance companies experience them. They also enable to explore the likelihood of future events and their financial impacts. The market consistent models provide a tool for valuing cash flows which depend upon stochastic financial variables. Although it is preferable that an ESG should address a broad-range of applications involving both real world and market consistent scenarios, one should have a comprehensive understanding of the specific applications and whether the applications require real world or market consistent scenarios, or both (Campbell et al., 2016).

1.4 One of the early ESGs which is specifically designed for actuarial applications is the Wilkie stochastic investment model (Wilkie, 1986; Wilkie, 1995; Wilkie et al., 2011). The Wilkie model is an open access ESG and it is the first comprehensive model designed to be used in the actuarial profession. The original Wilkie model was developed from the UK data over the period 1919-1982, and was made up of four interconnected models for price inflation, share dividend yields, share dividends and long-term interest rates. Wilkie (1995) updated the original model and extended it to include an alternative autoregressive conditional heteroscedastic (ARCH) model for price inflation, and models for wage inflation, short term interest rates, property yields and income and index-linked yields. Furthermore, these models were fitted to data from several developed countries and an exchange rate model was proposed. The model was updated in Şahin et al. (2008), Wilkie et al. (2011) and several futures of the stochastic investment models as well as additional economic data (earnings) have been analysed in the series of papers Wilkie & Şahin (2016, 2017a, 2017b, 2017c, 2018, 2019).

1.5 There is a well-developed literature on the real world ESGs, particularly discussing and criticising the Wilkie model as well as introducing similar models for different countries. Especially in the following ten years after its publication many other stochastic investment models were developed in a variety of ways. In this introduction, we consider mainly the real world ESGs developed for different countries such as Finland, U.K., Australia, Ghana, U.S. as well as paying specific attention to the South African literature.

1.6 Ranne (1998) proposed a stochastic investment model based primarily on Finnish data and different from the Wilkie model, he inserted a variable representing the economic cycles derived from the real growth rate of the gross national product. Yakoubov et al. (1999) introduced a stochastic investment model which is the first fully published model to use earnings rather than dividends to generate price returns. Whitten and Thomas (1999) suggested a non-linear model to introduce threshold modelling to the actuarial profession and illustrate how this can complement or replace methods based on autoregressive conditional heteroscedasticity. Chan (2002) adopted the multiple time-series modelling approach, a general VARMA (vector autoregressive moving average) model for the outlier adjusted UK investment data. The model is recommended for actuarial applications not involving extreme stochastic fluctuations. Sherris and Zhang (2009) proposed a multivariate regime switching vector autoregressive model which is calibrated to Australian data for the period 1979-2006. 11 economic variables representing main financial and economic series have been considered. They explore VAR (vector autoregressive), univariate and multivariate regime switching models and illustrate the relative performance of the models using simulation. It is concluded that the regime switching approach incorporated into the VAR model structure displays better performance when compared to unconditional historical distributions of the data series used in the model.

1.7 Beside those various ESGs, there are several researches on comparison of these types of models such as Harris (1995), Huber and Verrall (1999), Lee and Wilkie (2000), Rambaruth (2003) and Nam (2004). All these papers follow different methods to compare the models including re-estimating the parameters on the same interval, applying some model validation tests (independence, normality, likelihood), stability of the model parameters, calculating contingency reserves for specific contracts and forecasting by simulation.

1.8 Two recent papers, Tee & Ofosu-Hene (2017) and Zhang et al. (2018) compare and discuss the performance of the models for Ghana and the US. Tee & Ofosu-Hene (2017) fit and compare three actuarial stochastic asset models, namely the Wilkie model, the Wilkie ARCH model and Whitten and Thomas (1999) model to Ghanaian economic data. The analysis indicate that the investment series used in the paper are non-stationary, the simulated values using the Wilkie model produce similar characteristics to the historical data, the Wilkie model has a relatively better parsimony and ready economic interpretation while the Whitten and Thomas model produces simulated values with minimal forecast error. On the other hand, Zhang et al. (2018) update the Wilkie's ESG to 2014 using U.S. data and examine the stationarity assumptions and parameters stability considering the structural breaks to analyse the performance of the model. They conclude that the inflation violates the stationarity assumption and the parameters are sensitive to the calibration period. The out-of-sample backtest based on 30 years indicates that the model tends to overestimate inflation due to the structural shift in inflation targeting policy in the early 1990s and underestimates total return on stocks due to the dot-com bubble in the 1990s while it performs relatively well for long-term interest rates. Compared to a generically constructed ESG, the Wilkie model generates a wider range of scenarios for inflation and long-term interest but a narrower range for stock returns. Noting that the Wilkie model might be under-representative of the risk in long-term stock investment, particularly for tails, it is quite suitable to use for pension plans since the pension assets are relatively passively invested, so the annual time step suffices.

1.9 As mentioned previously, we focus on the South African stochastic investment models in this paper since the aim is to construct an updated stochastic investment model based on South African data to be used in long-term forecasting of economic variables for actuarial use.

1.10 The first comprehensive stochastic investment model for South Africa was introduced by Thomson (1996). The series modelled by Thomson were price inflation, short-term and long-term interest rates, dividend growth rates, dividend yields, rental growth rates and rental yields. No exogenous variables were included as in the Wilkie model, and the model was intended to be used in asset-liability modelling of South African defined benefit pension funds. Unlike Wilkie's model, Thomson's model was designed for projections of not more than ten years as having much shorter years of data available for South Africa (1960-1993). Due to the stationarity condition to apply Box & Jenkins methodology (Box and Jenkins, 1976), Thomson used prewhitened variables for his modelling work. Although it has a cascade structure, the order of the influence is different from the one in the Wilkie model. Thomson (1996) expresses his reservations about the validity of the model paucity of the data and he emphasises that it would be necessary to modify the model as time passes.

1.11 Thomson & Gott (2009) developed a long-term equilibrium model for South Africa which is different from the above descriptive stochastic investment models. They emphasise that the issues of arbitrage and equilibrium are generally not addressed and the models tend to be based on ex-post estimates. They argue that these descriptive models might produce risk-adjusted expected returns that exceed those of the market for some asset categories and understate those of the market for others. Therefore, they proposed an equilibrium model based on the returns on risk-free zero-coupon bonds both index-linked and conventional and on equities, as well as the inflation rate. The model is developed in discrete (nominally annual) time with an allowance for processes in continuous time subject to continuous rebalancing. The model is used as the basis of development of the arbitrage-free equilibrium model of its constituent asset categories.

1.12 Research in the area of stochastic models for actuarial use in SA is limited. Maitland (1996) reviews the Thomson model from a statistical and economic perspective. The paper was the first to do so and is largely critical of the Thomson model (developed as a forecasting tool) and concludes that the model should not be used in practical applications such as projections. The paper cited problems with the model structure identification as well as the method of estimating model parameters.

1.13 Maitland (1997) considers various descriptive models for variables such as inflation, equity dividend yields and dividend growth rates. The models presented are descriptive only and hence cannot be applied for projections. Maitland (2000) provides support for modelling the South African long-bond yield and the short rate as proposed by Thomson (1996). The paper provides a methodology for constructing the South African yield curve from these modelled yields using the first and second principal components.

1.14 Maitland (2010) presents a new stochastic model for South African equities, long and short-term interest rates and inflation. He recommends modelling the total return on equities instead of modelling equity dividend yields and growth rates separately. He recommends a regime switching model and introduces a Markov switching framework. Parameters of the model are estimated using historical data. The model is again a descriptive model (but does incorporate theoretical considerations).

1.15 In this paper, we introduce an updated "real world" stochastic investment model for long-term actuarial applications for South African use. Thomson (2013) discusses the assumption of "ergodicity" which states that the time series estimates serve as the unbiased estimators of the considered parameters. On this basis, we can categorise our model as an ESG based on the assumption of ergodicity. We use the phrase "real world" model for our economic scenario generator with some reservations since it is both based on the data and the implicit theory of the Wilkie Model. Although we referred to the Wilkie model in each economic series considered in our model, we tried to find the best model for the available data. Therefore, some of the economic variables produced different model structures than the Wilkie model. We consider the interaction between different economic series namely price inflation, share prices, share dividends, share dividend yields, long-term and short-term interest rates and inflation-linked bonds.

1.16 Following this introduction, the paper is structured as follows. Section 2 explains the methodology applied and the structure of the model. Sections 3 to 8 introduce the models, parameter estimates, model validation analysis and examine the parameter stability. Section 9 and 10 present the backtesting analysis and a brief insight for the practical implementations. Finally, Section 11 concludes the paper.

## 2. METHODOLOGY AND THE STRUCTURE OF THE MODEL

2.1 The model proposed in this paper is based on Box-Jenkins (1976) time series models. The parameters are estimated by using maximum likelihood method calculated by a non-linear optimisation method, the Nelder-Mead simplex method by using R programming language. All economic variables used in the model are stationary based on Kwiatkowski–Phillips–Schmidt–Shin (KPSS) test (1992) except dividend yields, which can be considered as stationary with 1% significance level. Some of the series are treated as if co-integrated, such as the logarithm of the share dividends and share prices as the difference gives the share dividend yields.

2.2 We use the annual June values for the period of 1960-2018 except for the inflation-linked bonds data which is available for 2000-2018. The main contribution of our paper is to develop an updated ESG based on South African data. Since we use 57-58 years of data for most of the economic variables considered in the model, our model can be used for long-term forecasting for actuarial purposes which cannot be done by the predecessor of this model introduced by Thomson (1996) due to the data restriction. This paper is the first to consider more than 50 years of data in SA to construct an ESG.

2.3 The fitted models for each economic series have been chosen among the candidates by applying some statistical tests on the residuals as model validation. The residuals have been analysed by calculating the auto-correlation functions, skewness and kurtosis as well as Jarque-Bera test to check if they are independent and identically distributed (iid) normal variates. The models producing higher log-likelihood values have been chosen over the ones which produce lower values.

2.4 The stochastic investment models which are designed for long-term applications should be analysed in terms of the stability of the parameters since the estimated parameters are assumed to be constant for the forecast period. We examine the parameter stability for each economic series separately by giving the details for price inflation model in Section 3 and presenting the results for the remaining series in Appendix A. As suggested by Huber (1997) and Wilkie et al. (2011), we investigate the parameter constancy of the models by recursively estimating the parameters on incrementally larger data sets and present 95% confidence intervals.

2.5 As a final model validation analysis, we fit the models chosen for the whole period (which ends 2018) for each economic series to the data ends with 2008 to estimate the parameters for a shorter period. Then, we use those parameters to forecast the remaining 10 years for each series and construct 95% and 99% forecast intervals to analyse the forecasting performance of our model. Although the model is designed for long-term applications, due to the data restrictions the backtesting is illustrated only for the last 10 years.

2.6 The series in the model are correlated and the model has a cascade structure. Figure 1 illustrates the structure of the model where the arrows indicate the direction of influence. Dotted arrows for share prices indicate that the share prices have not been modelled directly but derived from the models of share dividends and share dividend yields. The dotted arrow from long-term interest rates towards the inflation-linked bond rates indicates a possible statistically significant effect although the model involving short-term interest rates can be preferred to derive the inflation-linked bond rates as explained in the relevant section. One can see from the figure that the complete model is wholly self-contained. The only inputs are the separate white noise series, and no exogenous variables are included.

2.7 As in the Wilkie model, the price inflation is the driving force by affecting the share dividends and dividend yields as well as the interest rates. We use subscripts to distinguish the series: $q$ for price inflation, $y$ for dividend yields, $d$ for dividends, $c$ for long-term bond yields, $b$ for short-term bond yields and $r$ for inflation-linked bond yields.

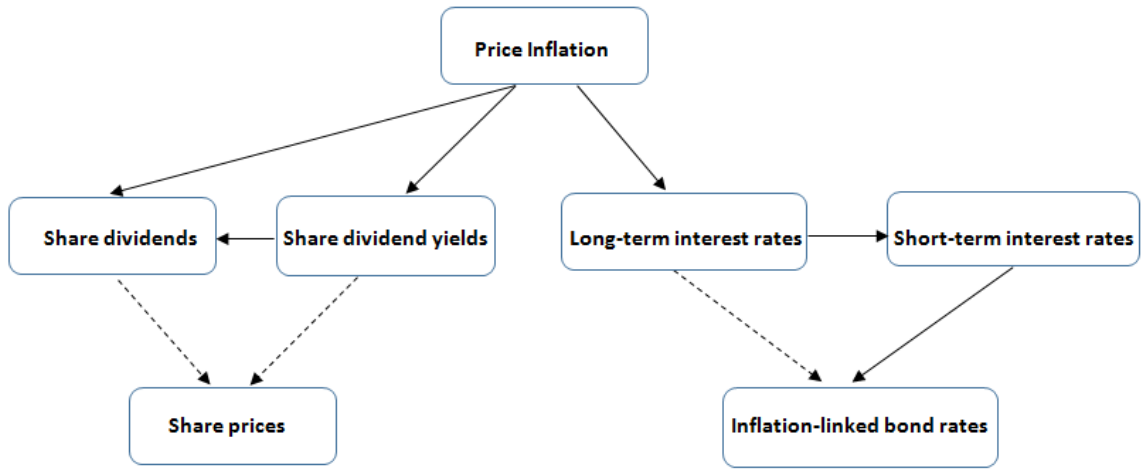

Figure 1. Structure of the model

## 3. PRICE INFLATION

3.1 To model the price inflation we use the South African Consumer Price Index (CPI) data which is provided by Statistics South Africa (2019). We model the force of inflation as a stationary autoregressive process. Denoting the value of CPI at time $t$ as $Q(t)$, we calculate the force of inflation over the year $(t-1,t)$, $\delta_q(t)$, as

$$Q(t) = Q(t-1).exp\left(\delta_q(t)\right), \qquad (1)$$

so that $\delta_q(t) = ln\{Q(t)\} - ln\{Q(t-1)\}$.

3.2 South Africa formally introduced inflation-targeting in February 2000. Since then, South Africa's Monetary Policy Committee has conducted its monetary policy to keep inflation within a target band of 3-6% (Monetary Policy Review, 2019). The reason for introducing the target ranges is to have a degree of flexibility for absorbing shocks outside the control of the authorities. The consumer price index, $Q(t)$, from 1959 to 2018 and the annual differences in the logarithms $\delta_q(t)$, are shown in Figure 2. Although it is not clear in Figure 2 whether the monetary policy is successful for the period 2004-2008 due to very low and high values, percentage changes in 12 months particularly after 2008 show that the average rate of increase in consumer prices, i.e. the bands together with the actual inflation rates, are achieved (South African Reserve Bank, 2019).

3.3 The force of inflation $\delta_q(t)$ which is defined as the difference in the logarithms of the CPI each year, is modelled as a first order autoregressive (AR) series as given in Eq. (2). An AR(1) model is a statistically stationary series for suitable parameters, which means that in the long run the mean and variance are constant. We define the model as below.

$$\delta_q(t) = \mu_q + a_q\left(\delta_q(t-1) - \mu_q\right) + \epsilon_q(t), \qquad (2)$$

$$\epsilon_q(t) = \sigma_q.z_q(t), \qquad (3)$$

$$z_q(t) \sim iidN(0,1), \qquad (4)$$

where $\mu_q$ is the long-run mean, $a_q$ is the autoregressive parameter, $\sigma_q$ is the standard deviation of the residuals and $z_q(t)$ is a series of independent, identically distributed unit normal variates. The model states that each year the force of inflation is equal to its mean rate, $\mu_q$, plus some proportion, $a_q$, of last year's deviation from the mean, plus a random innovation which has zero mean and a constant standard deviation, $\sigma_q$.

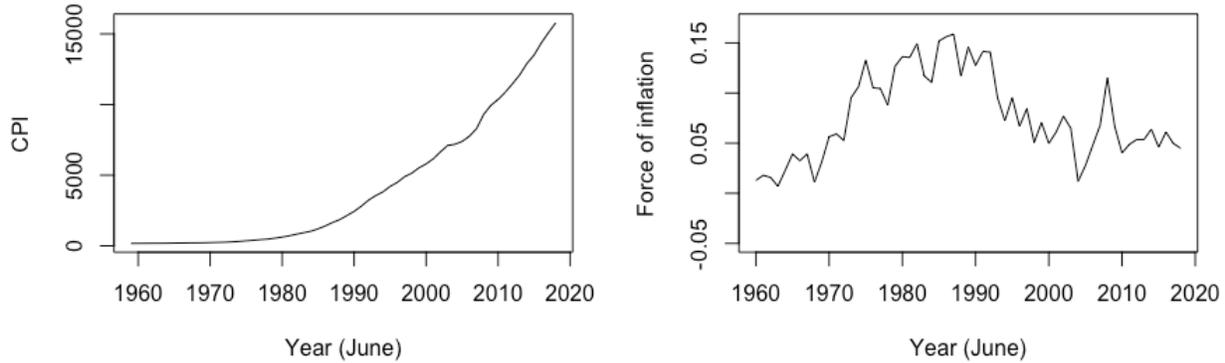

Figure 2. Consumer Price Index (CPI), 1959-2018

3.5 The model validation has been analysed by applying some statistical tests on the residuals. The autocorrelation coefficients of the residuals, $r_z$, and squared residuals, $r_z^2$, show nothing unusual, i.e. residuals can be considered to be independent and there is no simple autoregressive conditional heteroskedasticity (ARCH) effect. The skewness and kurtosis coefficients, based on the third and fourth moments of the residuals, are close to the theoretical values of the normal distribution (zero for skewness and 3 for kurtosis). A composite test of the skewness and kurtosis coefficients, Jarque-Bera test statistic is 0.2191 for the observation period, which should be compared with a $\chi^2$ variate with two degrees of freedom. The p-value is 0.90 and therefore, the normality assumption holds.

**3.6 Parameter Stability**

3.6.1 As mentioned in Section 2 we investigate the parameter constancy of the models by recursively estimating the parameters on incrementally larger data sets. Figure 3 presents these recursive estimates and 95% confidence intervals of $\mu_q$, $a_q$ and $\sigma_q$, for *earlier sub-periods* (data sets starting in 1985) and *later sub-periods* (data sets ending in 2018) for the price inflation model. Sub-periods with fewer than 25 observations for the early periods and 10 observations for the later periods are omitted in this case to obtain reasonable parameter values.

3.6.2 Solid lines show the parameter estimates and the dotted lines show the 95% confidence intervals in Figure 3. These are based on an assumption that the parameter value is distributed normally, and are calculated as the estimated value plus or minus 1.96 times the calculated standard error.

3.6.3 The heavy lines show the estimated values of $\mu_q$, $a_q$ and $\sigma_q$ for periods starting in 1960 and ending in the given year. It begins with the period ending in 1985, for which there are 25 observations from which to estimate the parameters. Over this period, we see that the estimated value of $\mu_q$ is 11.03% and increases up to 16.17% in the next four years. Then, it steadily decreases with small distortions and ends with 8.09% (the long-term mean) in 2018.

| Inflation Model | |
|---|---|
| (1960-2018) | |
| $\delta_q(t)$ | AR(1) |
| $\mu_q$ | 0.0809 (0.0185) |
| $a_q$ | 0.8433 (0.0670) |
| $\sigma_q$ | 0.0220 (0.0020) |
| Log Likelihood | 139.05 |
| $r_z(1)$ | -0.119 |
| $r_{z^2}(1)$ | -0.043 |
| skewness $\sqrt{\beta_1}$ | -0.1031 |
| kurtosis $\beta_2$ | 2.7841 |
| Jarque-Bera $\chi^2$ | 0.2191 |
| $p(\chi^2)$ | 0.8962 |

Table 1. Estimates of parameters and standard errors (in brackets) of AR(1) model for inflation over 1960-2018

3.6.4 The thinner continuous line in the graphs show the estimated values of $\mu_q$, $a_q$ and $\sigma_q$ for periods ending in 2018. These lines commence in 1960 at the value 8.09% for $\mu_q$, being the value for the whole period 1960-2018. The line lowers slightly ending at 5.78% in 2009, the last year for which we have 10 years data ending in 2018. The estimated mean value for the last 10 years (5.78%) being in the targeted band (3-6%) indicates that the inflation-targeting policy seems to work.

3.6.5 The dotted lines of either side of the thinner continuous line show approximate 95% confidence intervals for the corresponding value; the dash lines on either side of the heavy line do the same.

3.6.6 Figure 3 shows that $\sigma_q$ values over the years are relatively stable considering the range of the parameter estimates while $\mu_q$ and $a_q$ parameters seem more sensitive to the different periods of data. Recursive estimates for $a_q$ for the later sub-periods and recursive estimates for $\mu_q$ for the earlier sub-periods are quite volatile which also coincide with the high and/or unstable values of price inflation for the periods 1980-1990 and 2000-2010.

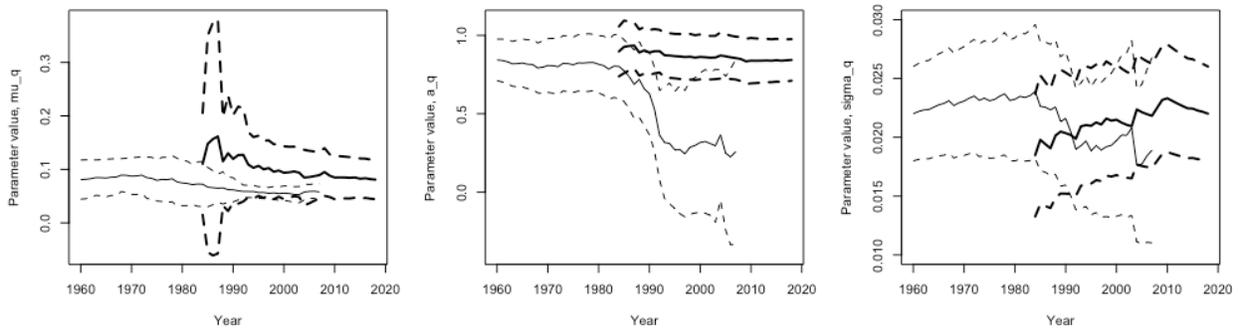

Figure 3. Estimates for parameters $\mu_q$, $a_q$ and $\sigma_q$ (solid lines show the parameter estimates and dotted lines show the 95% confidence intervals)

## 4. SHARE DIVIDEND YIELDS

4.1 We use the dividend yield on the JSE-Actuaries All Share Index (ADY) provided until June 2001 which was replaced by the ALSI Dividend Yield (J202) afterwards to construct a model for share dividend yields.

4.2 Figure 4 shows the share dividend yields in percentages which decrease after mid 1980s and shows some significant jumps.

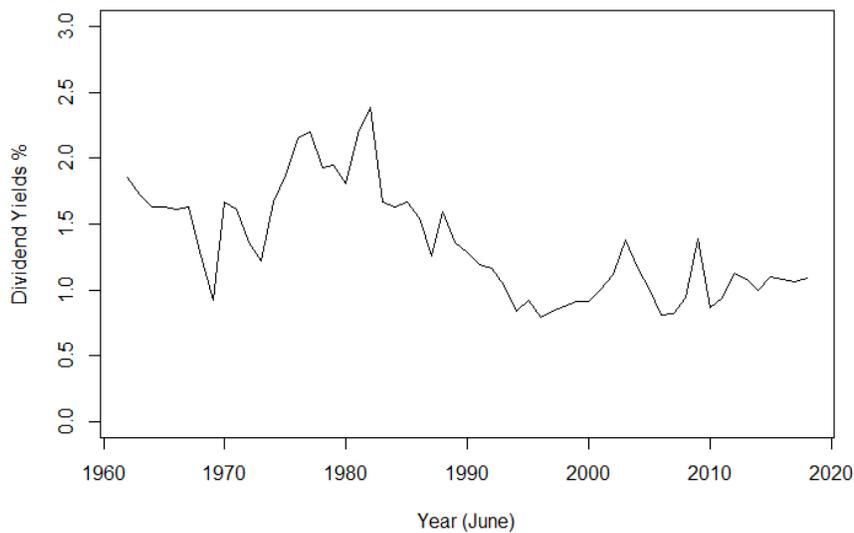

Figure 4. Dividend Yields %, 1961-2018

4.3 Although we fitted several models to the dividend yields, we ended up including both autoregressive and inflation effects to reach a satisfactory model given in Eqs. 5-10 as below. $y(t)$ is the yield on the index at time $t$:

$$ym(t) = d_y.\delta_q(t) + (1 - d_y).ym(t-1), \qquad (5)$$

$$y_q(t) = w_y.ym(t) + (1 - w_y).\delta_q(t), \qquad (6)$$

$$\ln y(t) = y_q(t) + \mu_y + yn(t), \qquad (7)$$

$$yn(t) = a_y.yn(t-1) + \varepsilon_y(t) \quad , \qquad (8)$$

$$\varepsilon_y(t) = \sigma_y.z_y(t), \qquad (9)$$

$$z_y(t) \sim iidN(0,1), \qquad (10)$$

where $ym(t)$ is the moving average effect of inflation which has an inflation component $\delta_q(t)$, $yn(t)$ is the autoregressive part and $z_y(t)$ is a series of independent, identically distributed unit normal variates.

4.4 We used MLE method to estimate the parameters of the models presented in Table 2. We compare two models - AR(1) and AR(1) with moving average (MA) inflation effect on dividend yields based on the log-likelihood values and the statistics which are used to check whether the model assumptions such as normality and independence of the residuals are held. Including the inflation effect improves the log-likelihood significantly which also justifies the two additional parameters. Jarque-Bera test indicates that having a model with inflation effect satisfies the necessary normality assumption.

| | **Dividend Yields Model** (1962-2018) | |
|---|---|---|
| $\delta_y(t)$ | AR(1) | MA Inflation effect |
| $w_y$ | | -4.0074 (1.2161) |
| $d_y$ | | 0.1396 (0.0557) |
| $\mu_y$ | 1.2695 (0.1774) | 0.3781 (0.1152) |
| $a_y$ | 0.8266 (0.0727) | 0.6318 (0.0890) |
| $\sigma_y$ | 0.2261 (0.0214) | 0.1973 (0.0186) |
| Log Likelihood | 3.81 | 11.42 |
| $r_z(1)$ | -0.049 | -0.132 |
| $r_{z^2}(1)$ | 0.180 | -0.036 |
| skewness $\sqrt{\beta_1}$ | 0.5254 | 0.2350 |
| kurtosis $\beta_2$ | 4.0364 | 2.9365 |
| Jarque-Bera $\chi^2$ | 5.1731 | 0.5343 |
| $p(\chi^2)$ | 0.0753 | 0.7656 |

Table 2. Estimates of parameters and standard errors (in brackets) of the models for dividend yields over 1962-2018

## 5. SHARE DIVIDENDS

5.1 Thomson (1996) used JSE-Actuaries All Share Index (CI101) for the purpose of his modelling. This was later replaced by the ALSI Total Return Index (J203). There was an overlap period from 1995-2002. Because of the discrepancies between the construction of the CI01 and J203 indices, their total returns figures did not reconcile in the period 1995-2002 when both were being calculated and published concurrently. In order to avoid a discontinuity in the index movements over time, the J203 index was rebased to match the value of the CI01 on the date the J203 came into effect. The total return is then calculated on the rebased J203 index as opposed to using the CI01 index. We use this rebased index for the share prices and obtain share dividends.

5.2 Share dividends, $D(t)$, are obtained from the published values of share prices, $P(t)$, and dividend yields, $Y(t)$ as $D(t) = P(t).Y(t)$.

5.3 The logarithm of the dividend growth, $\delta_d(t) = lnD(t) - lnD(t-1)$ is plotted in Figure 5. Although there are some negative and positive jumps in the yearly dividend values, the overall level of the index does not change significantly and the data seems stationary.

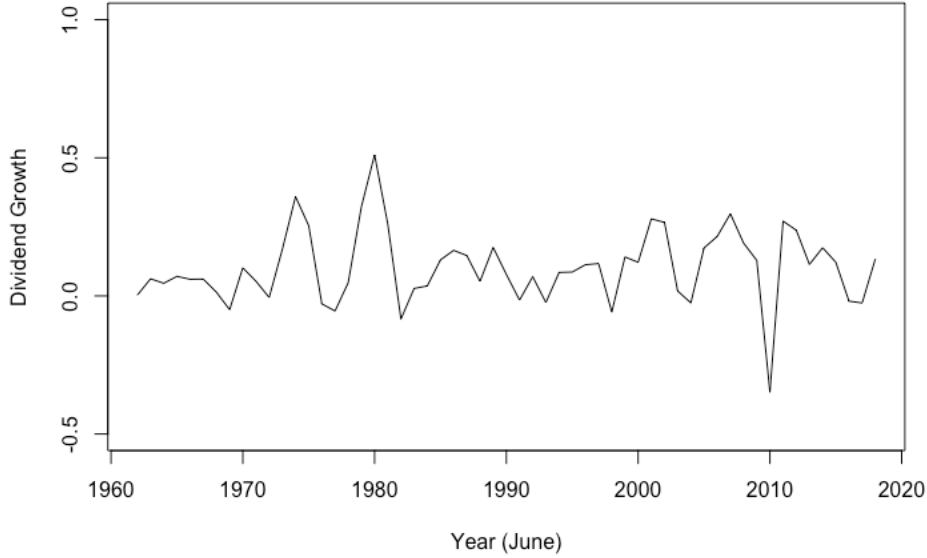

Figure 5. Share Dividends Growth, 1962-2018

5.4 The logarithm of the dividend growth, $\delta_d(t)$, can be modelled including the effect from previous year's dividend yields residuals $\varepsilon_y(t-1)$ as well as a simultaneous inflation effect $\delta_q(t)$, or a moving average inflation effect, $dm(t)$ where $z_d(t)$ is a series of independent identically distributed unit normal variates as presented in Eqs. 11-15. $dm(t)$ is a function of exponentially weighted moving average of inflation up to time $t$. As in the Wilkie model, $d_q(t)$ takes a proportion of the moving average inflation effect and a proportion of the latest rate of inflation and the coefficients $w_q$ and $1 - w_q$ indicate that there is 'unit gain' from inflation to dividends. $\delta_d(t)$ is also influenced by the residuals from the previous year of dividend yields and the dividend growth itself.

$$dm(t) = d_d.\delta_q(t) + (1 - d_d).dm(t-1), \tag{11}$$

$$d_q(t) = w_d.dm(t) + (1 - w_d)\delta_q(t), \tag{12}$$

$$\delta_d(t) = d_q(t) + \mu_d + y_d.\varepsilon_y(t-1) + k_d.\varepsilon_d(t-1) + \varepsilon_d(t), \tag{13}$$

$$\varepsilon_d(t) = \sigma_d.z_d(t), \tag{14}$$

$$z_d(t) \sim iidN(0,1). \tag{15}$$

**Share Dividends Model**

**(1962-2018)**

| $\delta_d(t)$ | only $\delta_y$ effect | +Inflation effect | +MA Inflation effect |
|---|---|---|---|
| $w_d$ | | | -5.5068 (3.6008) |
| $d_d$ | | | 0.6499 (0.1970) |
| $q_d$ | | 0.9482 (0.4815) | |
| $\mu_d$ | 0.1415 (0.0276) | 0.0675 (0.0460) | 0.0649 (0.0245) |
| $y_d$ | -0.1622 (0.0728) | -0.1673 (0.0692) | -0.1850 (0.0690) |
| $k_d$ | 0.3670 (0.1292) | 0.3828 (0.1272) | 0.2798 (0.1479) |
| $\sigma_d$ | 0.1207 (0.0114) | 0.1166 (0.0110) | 0.1086 (0.0103) |
| Log Likelihood | 38.97 | 40.89 | 44.87 |
| $r_z(1)$ | -0.023 | -0.011 | -0.005 |
| $r_{z^2}(1)$ | 0.198 | 0.198 | 0.103 |
| skewness $\sqrt{\beta_1}$ | 0.0266 | -0.0589 | -0.0360 |
| kurtosis $\beta_2$ | 4.2681 | 3.8662 | 3.3556 |
| Jarque-Bera $\chi^2$ | 3.8259 | 1.8151 | 0.3125 |
| $p(\chi^2)$ | 0.1476 | 0.4035 | 0.8553 |

Table 3. Estimates of parameters and standard errors (in brackets) of the models for share dividends over 1962-2018

5.5 Table 3 presents the parameter values, standard errors, log-likelihood values and the statistical tests for each model. All three models seem satisfactory while the one with both dividend yields and moving average inflation effect seems the best based on the log likelihood and model validation tests. Using the models for share dividend yields and share dividends it is straightforward to obtain share prices.

## 6. LONG-TERM INTEREST RATES

6.1 For long-term interest rates, the JSE-Actuaries Long Bond Yield (i.e. JAYC20, the 20-year Bond Yield) is used as in Thomson (1996). The data is provided by IRESS (2019).

6.2 Figure 6 shows that long-term and short-term interest rates move closely and they are also correlated with price inflation. The autocorrelation and cross correlation functions (CCF) indicate significant correlations between long-term interest rates and historical price inflation going back 5 to 10 years. Additionally, long-term interest rates and dividend yield residuals have simultaneous and lagged correlations. We investigate whether those correlations can be incorporated to obtain a more sophisticated and economically meaningful long-term interest rates model. Our analysis showed that the parameter which represents the dividend yield residuals is not significant so we eliminate the dividend yield effect.

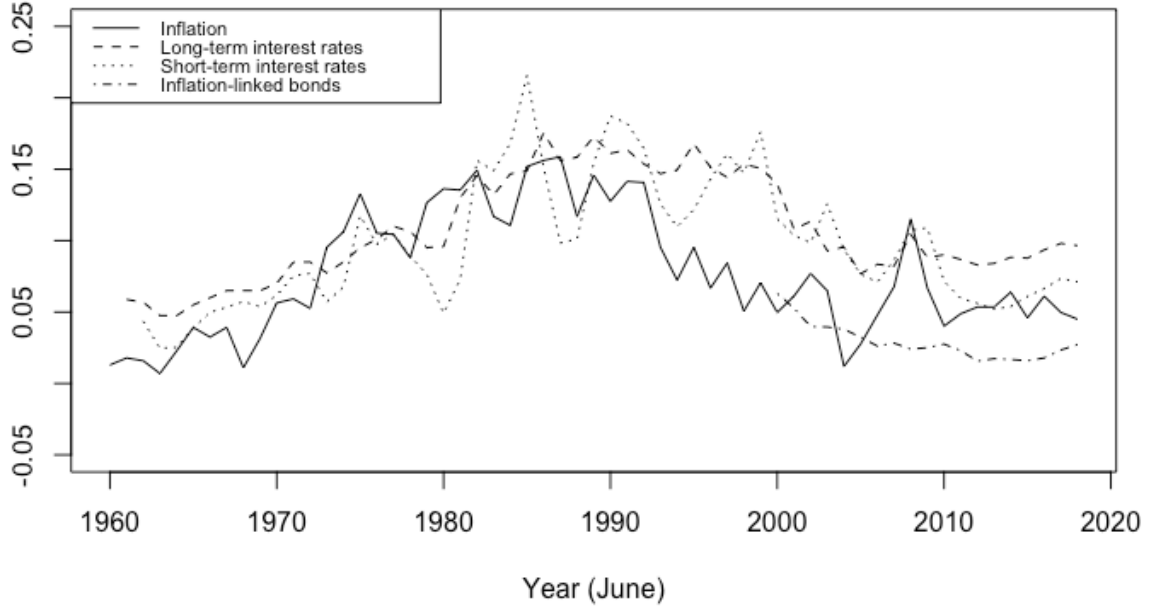

Figure 6. Annual force of inflation based on CPI, long-term interest rates, short-term interest rates and inflation-linked bond rates, 1960-2018

6.3 We denote long-term interest rates as $\delta_c(t)$, where $cm(t)$ represents the inflation effect, $cr(t)$ is the real interest rates obtained as the difference between the nominal rates and the inflation, $cn(t)$ is the autoregressive effect and $z_c(t)$ is a series of independent identically distributed unit normal variates.

6.4 Wilkie et al. (2011) fixed the parameters $w_c = 1$ $(CW = 1)$ and $d_c = 0.045 (CD = 0.045)$ in the consols yield model to ensure that the real interest rates were never negative for the period considered in the model. We inserted the moving average inflation effect by fixing $w_c = 1$ and $d_c = 0.13$ after several trials and the formulas are given in Eqs (16)-(21) below.

$$cm(t) = d_c. \delta_q(t) + (1 - d_c). cm(t-1), \qquad (16)$$

$$cr(t) = \delta_c(t) - w_c. cm(t), \qquad (17)$$

$$\ln cr(t) = \ln \mu_c + cn(t), \qquad (18)$$

$$cn(t) = a_c. cn(t-1) + \epsilon_c(t), \qquad (19)$$

$$\epsilon_c(t) = \sigma_c. z_c(t), \qquad (20)$$

$$z_c(t) \sim iidN(0,1). \tag{21}$$

6.5 Table 4 presents the parameters for the AR(1) model and the model with moving average inflation effect. Although the cross correlations between the long-term interest rates and the inflation indicate strong simultaneous and lagged correlations in Figure 6, inserting relevant parameters worsens the fit of the model compared to the AR(1) model. We present both models and estimate the parameter stability based on the second model (results are given in Appendix B) since we believe that it is important to consider the economic theory as well as the data and statistics in stochastic investment modelling. Figure 7 shows the cross correlations between the interest rates and inflation, Fisher relation based on the estimated moving average inflation effect and the real interest rates derived from the model and Efficient Market Hypothesis which shows the close relation between the inflation effect captured in the model and the realised inflation over the years.

| | **Long-term Interest Rates Model** | |
| --- | --- | --- |
| | **(1961-2018)** | |
| $\delta_c(t)$ | AR(1) | MA Inflation effect |
| $w_c$ | | 1.0 |
| $d_c$ | | 0.13 |
| $\mu_c / \ln \mu_c$ | 0.1174 (0.0235) | -3.3892 (0.1086) |
| $a_c$ | 0.9328 (0.0418) | 0.5665 (0.1117) |
| $\sigma_c$ | 0.0115 (0.0010) | 0.3610 (0.0341) |
| Log Likelihood | 173.59 | 29.06 |
| $r_z(1)$ | -0.101 | -0.108 |
| $r_{z^2}(1)$ | -0.016 | 0.109 |
| skewness $\sqrt{\beta_1}$ | 0.3227 | -0.9368 |
| kurtosis $\beta_2$ | 3.9470 | 4.2522 |
| Jarque-Bera $\chi^2$ | 3.17 | 12.06 |
| $p(\chi^2)$ | 0.2046 | 0.0024 |

Table 4. Estimates of parameters and standard errors (in brackets) of the model for long-term interest rates over 1961-2018

6.6 The Fisher relation (Fisher, 1930) states that expected inflation is fully reflected in nominal interest rates. As a result, this relation assumes that investors' expectations of average future inflation can be approximately determined by subtracting the average future real return required by investors from nominal interest rates. As in the Wilkie model, the Fisher relation is explicitly included in the long-term interest rate model (+MA inflation effect) by assuming that the average future real return required by investors is given by $cr(t)$ and that investors' expectation of average future inflation is given by $cm(t)$. Based on our long-term interest rate model, average expected future inflation is 7.23% and average future real return is 3.60%. These percentages are consistent with the average realised inflation of South Africa over the period 1961-2018 which is 7.79%. Due to the lack of real interest rate historical data we can compare the $cr(t)$ values over the period of 2000-2018 with the inflation-linked bond rates which will be discussed in the following section. The average real interest rate is 2.58% based on the Bloomberg's inflation-linked bond prices while the average future real returns is equal to 2.91% obtained from our long-term interest rates model. These averages show that although inserting the inflation effect to the long-term interest rates model worsens the fit, the extended model provide consistent information regarding the future average inflation rate and the investors' expectations for the future real returns.

6.7 Figure 7 also presents the graph for the rational expectation hypothesis. The concept of rational expectations asserts that outcomes do not differ systematically (i.e., regularly or predictably) from what people expected them to be. It does not deny that people often make forecasting errors, but it does suggest that errors will not persistently occur on one side or the other. Although the future is not fully predictable, agents' expectations are assumed not to be systematically biased. The third graph in Figure 7 shows the realised inflation $\delta_q(t)$, smoothed expected inflation obtained from the long-term interest rates model, $cm(t)$ with the optimal estimate of average future inflation which is equal to $\mu_q$. Comparing those three inflation information we can see that up to 1995 the investors slightly underestimate the average future inflation while they slightly overestimate it afterwards. The differences between the estimated and realised average inflation values are not high and there is no systematic pattern in the differences.

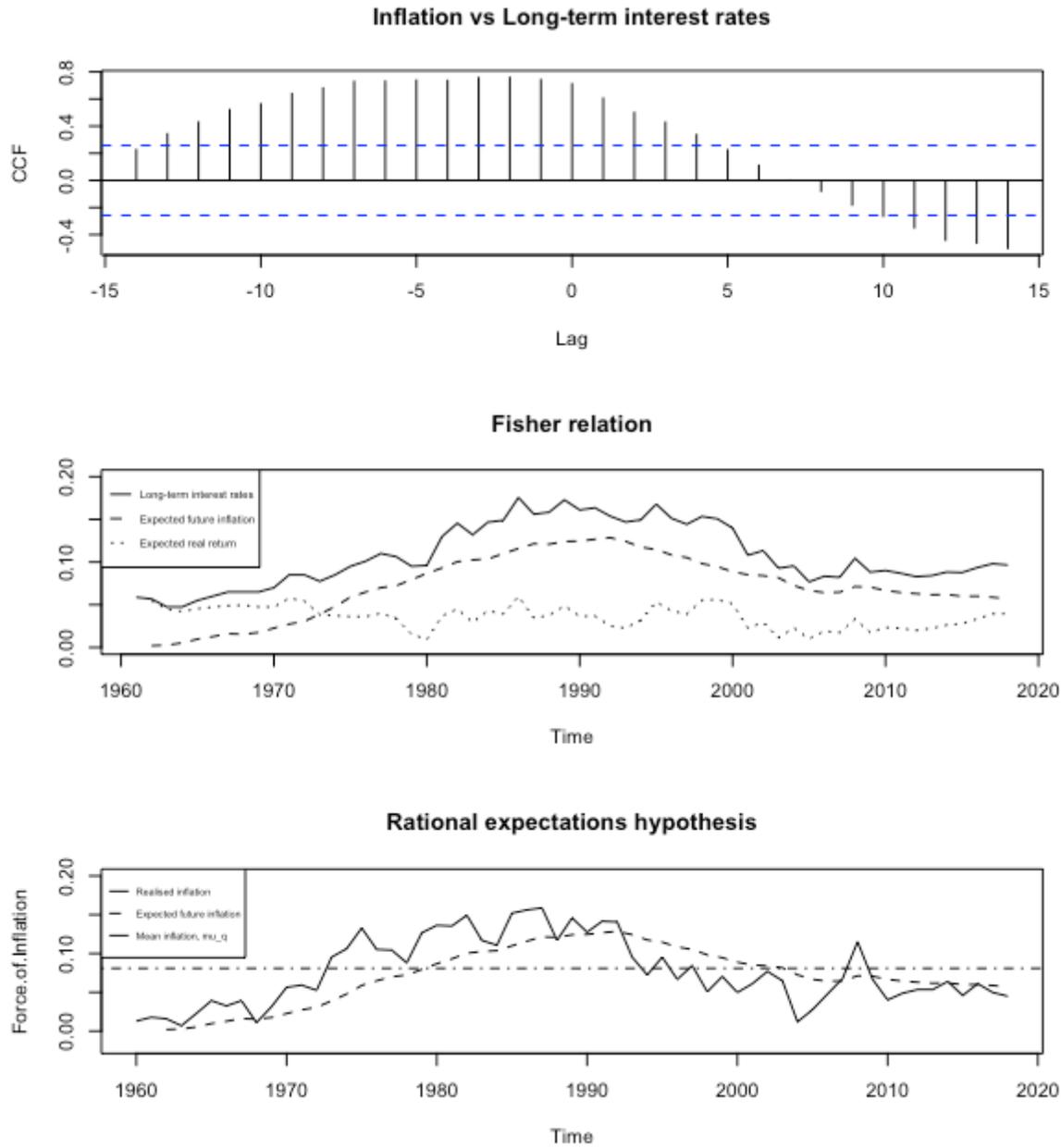

Figure 7. Cross correlation function (CCF) of inflation and long-term interest rates, Fisher relation and Efficient market hypothesis

## 7. SHORT-TERM INTEREST RATES

7.1 As in Thomson (1996), the Alexander Forbes Money-Market Index (GMC1) (previously referred to as the Ginsburg Malan & Carsons Money-Market Index) has been used for the short-term interest rates model and the data is provided by IRESS (2019). The short-term interest rates, $\delta_b(t)$, is obtained in Eq (22) as

$$\delta_b(t) = ln\frac{GMC1(t)}{GMC1(t-1)}, \qquad (22)$$

based on the data.

7.2 Figure 8 shows that short-term interest rates are clearly connected with long-term ones. Wilkie's approach was to model the difference between the logarithms of these series where $\delta_c(t)$ is the long-term interest rates and $\delta_b(t)$ is the short-term interest rates:

$$ln\delta_c(t) - ln\delta_b(t) = -ln\frac{\delta_b(t)}{\delta_c(t)} \qquad (23)$$

i.e. the logarithm of the ratio of the rates.

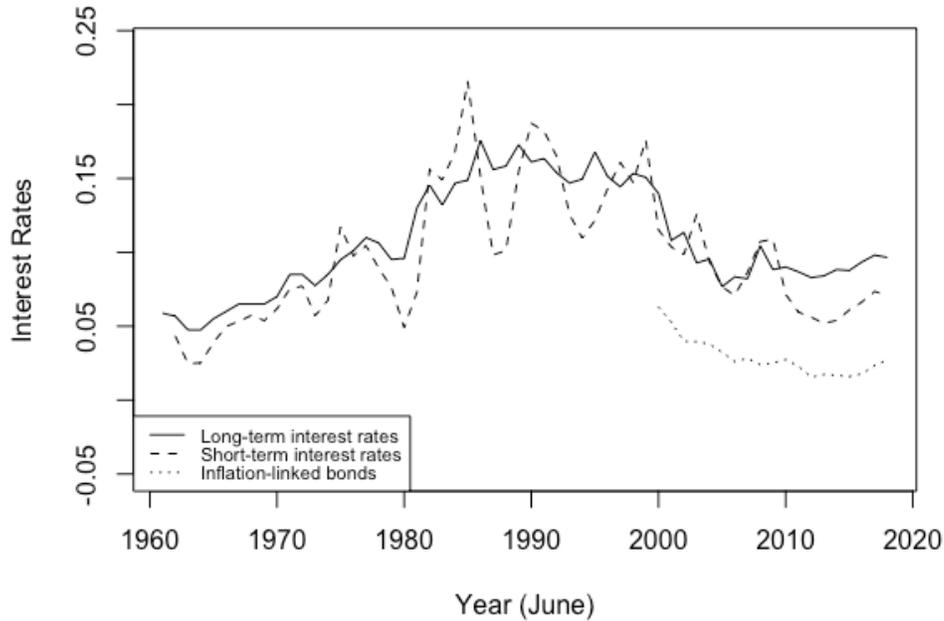

Figure 8. Long-term Interest Rates, Short-term Interest Rates and Inflation-linked Bonds, 1961-2018

7.3 We follow Wilkie's approach and the inspection of the data shows that AR(1) model fits the log ratio, $bd(t)$ quite well. The short-term rate of interest at time $t$ is defined as $\delta_b(t)$ and we put:

$$\delta_b(t) = \delta_c(t).exp\,(-bd(t)), \qquad (24)$$

where:

$$bd(t) = \mu_b + a_b.(bd(t-1) - \mu_b) + \epsilon_b(t), \qquad (25)$$

$$\epsilon_b(t) = \sigma_b.z_b(t), \qquad (26)$$

$$z_b(t) \sim iidN(0,1). \qquad (27)$$

Note that $bd$ has a minus sign in front of it, because short-term yields are, on average, lower than long-term ones. Figure 9 presents the log spread values for the period 1963-2018.

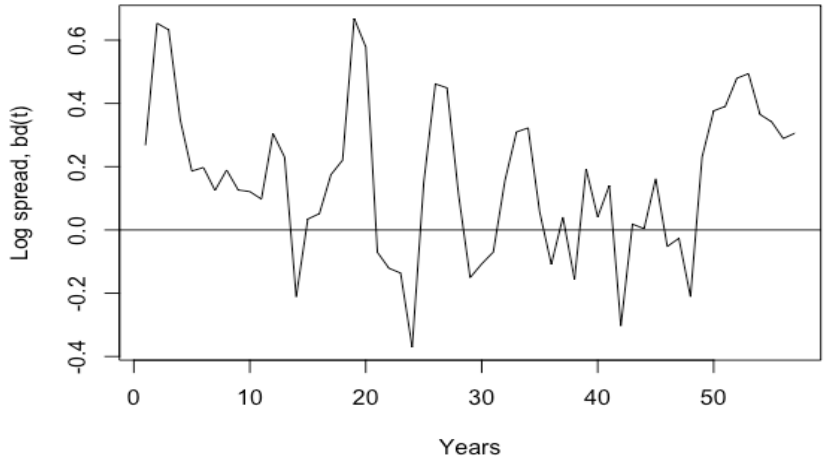

Figure 9. Log spread, $bd(t) = ln(\delta_c(t)/\delta_b(t))$, 1963-2018

7.4 Table 5 presents the parameter values, standard errors and the statistics based on the AR(1) model fitted to the log ratio. The estimated parameters are significant, the residuals are distributed normal and the model satisfies the necessary independence and normality assumptions.

| Short-term Interest Rates Model | |
|---|---|
| **(1962-2018)** | |
| $bd(t)$ | AR(1) |
| $\mu_b$ | 0.1568 (0.0596) |
| $a_b$ | 0.5527 (0.1116) |
| $\sigma_b$ | 0.1996 (0.0189) |
| $r_z(1)$ | -0.095 |
| $r_{z^2}(1)$ | 0.098 |
| skewness $\sqrt{\beta_1}$ | -0.2012 |
| kurtosis $\beta_2$ | 3.1408 |
| Jarque-Bera $\chi^2$ | 0.4318 |
| $p(\chi^2)$ | 0.8058 |

Table 5. Estimates of parameters and standard errors (in brackets) of the model for short-term interest rates over 1962-2018

## 8. INFLATION-LINKED BONDS

8.1 The inflation-linked bonds data is obtained from Bloomberg (2019). We use an arithmetic average of the real yields of the government inflation-linked bonds in issuance for different maturities due to lack of historical data for a specific maturity. The first issuance was in March 2000 as shown in Figure 10.

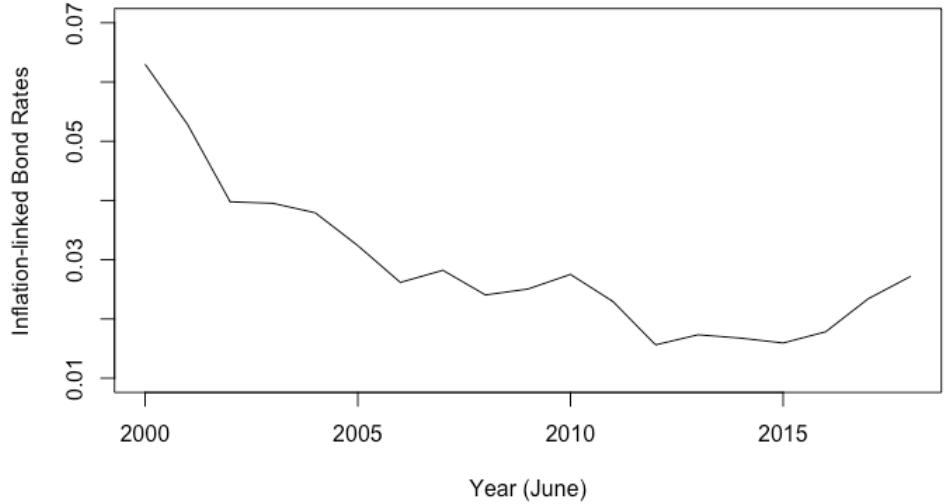

Figure 10. Inflation-linked bond rates, 2000-2018

8.2 We investigated the cross correlations between inflation, long-term and short-term interest rates and the inflation-linked bonds to construct a model for the real interest rates. We have fitted many different models by considering different relations between the series as presented in Table 6. The most comprehensive model is given in Eqs. (28)-(30) where $\delta_r(t)$ is the real interest rate, $\delta_c(t)$ is the long-term interest rate and $\delta_b(t)$ is the short-term interest rate, $\mu_r$ is the overall mean, $a_r$ is the autoregressive parameter, $c_r$ and $b_r$ are the long-term and short-term interest rates parameters and finally $z_r(t)$ is a series of independent identically distributed unit normal variates.

$$\delta_r(t) = \mu_r + a_r.(\delta_r(t-1) - \mu_r) + c_r.\delta_c(t) + b_r.\delta_b(t) + \epsilon_r(t), \quad (28)$$

$$\epsilon_r(t) = \sigma_r.z_r(t), \quad (29)$$

$$z_r(t) \sim iidN(0,1). \quad (30)$$

Table 6 shows that all models satisfy the necessary assumptions while there are slight differences between their goodness of fits and log-likelihood values. The simple AR(1) model and the model incorporating the short-term interest rates effect are the best models for the period under consideration.

| $\delta_r(t)$ | AR(1) | $\delta_c(t)$ and $\delta_b(t)$ | $\delta_c(t)$ | $\delta_b(t)$ | |
|---|---|---|---|---|---|
| | | | | $\mu_r$ included | $\mu_r$ omitted |
| $\mu_r$ | 0.0222 (0.0033) | 0.0438 (0.0316) | -0.0118 (0.0266) | 0.0038 (0.0071) | |
| $a_r$ | 0.7194 (0.0618) | 0.5942 (0.0957) | 0.6877 (0.0745) | 0.6174 (0.0698) | 0.6165 (0.0703) |
| $c_r$ | | -0.2721 (0.1582) | 0.1163 (0.0889) | | |
| $b_r$ | | 0.2142 (0.0727) | | 0.0973 (0.0419) | 0.1144 (0.0272) |
| $\sigma_r$ | 0.0033 (0.0004) | 0.0038 (0.0007) | 0.0034 (0.0004) | 0.0029 (0.0003) | 0.0030 (0.0003) |
| Log Likelihood | 77.10 | 74.61 | 76.95 | 79.46 | 79.32 |
| $r_z(1)$ | 0.038 | | | -0.068 | -0.060 |
| $r_{z^2}(1)$ | -0.072 | | | -0.230 | -0.151 |
| skewness $\sqrt{\beta_1}$ | -0.2749 | | | -0.3155 | -0.3418 |
| kurtosis $\beta_2$ | 2.2125 | | | 2.2151 | 2.3209 |
| Jarque-Bera $\chi^2$ | 0.7303 | | | 0.8030 | 0.7349 |
| $p(\chi^2)$ | 0.6941 | | | 0.6693 | 0.6925 |

**Inflation-linked Bond Yields (2000-2018)**

Table 6. Estimates of parameters and standard errors (in brackets) of the model for inflation-linked bond rates over 2000-2018

8.3 Due to having a very short period of data, we cannot examine the stability of the parameters for this model.

## 9. BACKTESTING

9.1 The performance of our proposed model is assessed in this section based on out-of-sample model validation. To backtest the model, we used the selected models for each economic series (except inflation-linked bond yields due to short period of data) and fit the models to data from 1960 to 2008 as the longest period to estimate the parameters. Then we use these parameters to project 100,000 scenarios of 10 years and display the 95% and 99% forecast intervals compared to the historical observations for the period 2008-2010 in Figure 11. Estimated parameters are given in Appendix B.

9.2 The graphs for funnels of doubt for the economic series show that almost all forecasts for the next 10 years are within the confidence intervals although the funnels of doubt are getting wider as time passes. The inflation model tends to overestimate due to a higher value of long-term mean parameter based on the period 1960-2008. The estimated mean parameter might need further adjustments to produce more realistic forecasts, which is briefly discussed in the following section. The forecast results presented below are all calculated by using the estimated parameters from the data up to 2008. CPI forecasts have been derived from the forecasted inflation rates.

9.3. The stationarity problem of the dividend yields data is reflected in the forecast intervals as producing quite wide ranges. It seems like the high values in the historical data dominate the parameter estimations and forecasts. The prediction interval for dividends is stable and narrow although the record low value observed in 2010 lies outside of the range.

9.4 The funnels of doubt increase as forecast period increases for long-term (both AR(1) and moving average inflation effect models) and short-term interest rates. The high interest rates between 1980 and 2000 cause higher estimated mean values which also affects the forecasts. Thus, the models produce interest rates above the observed rates for the period 2008-2018 although all lie within the forecast intervals. The short-term interest rates model requires forecasted values of the long-term interest rates which can explain the wider funnel of doubt even for the early years of forecasts.

9.5 The parameter stationarity analysis of the models given in Appendix A indicate time-dependence of some parameters such as inflation effect parameters of dividend yields and share dividends and mean parameter of the long-term interest rates. The higher number of parameters to be estimated in a model fitted to a short period of data also cause non-stable estimates. As presented in Appendix B, the parameters estimated for the periods up to 2008 have slightly higher standard errors for almost all models compared to the ones estimated for the whole period.

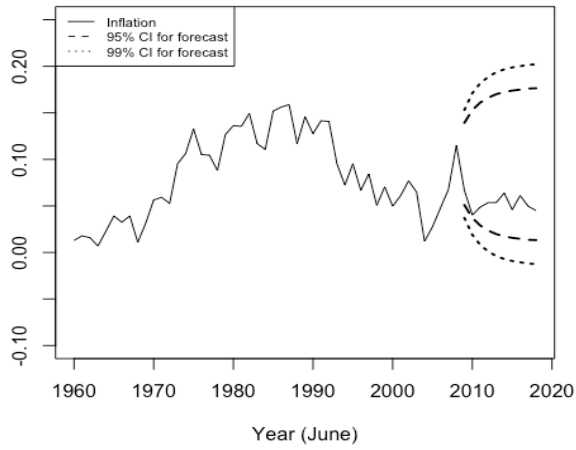
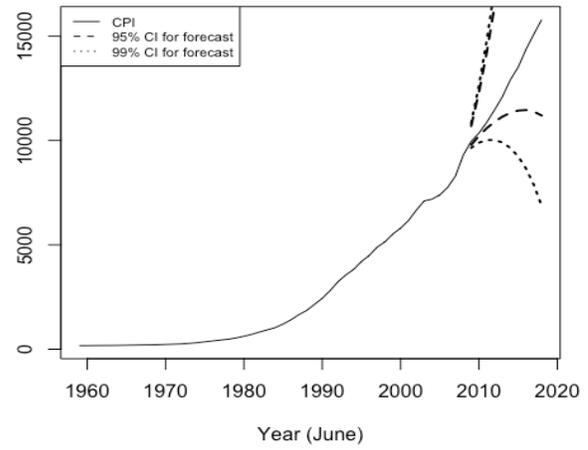
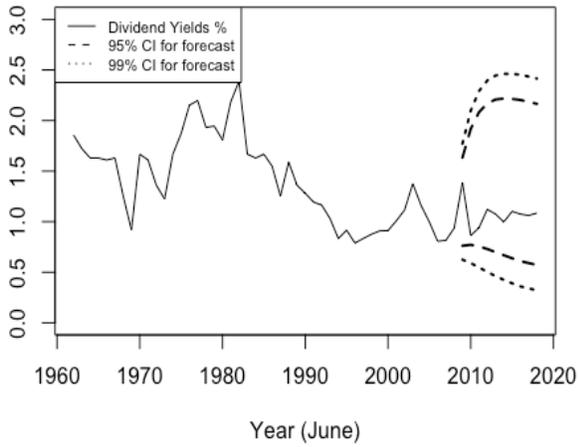
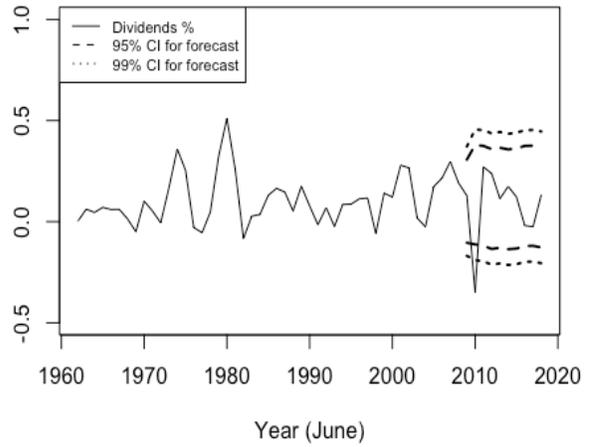
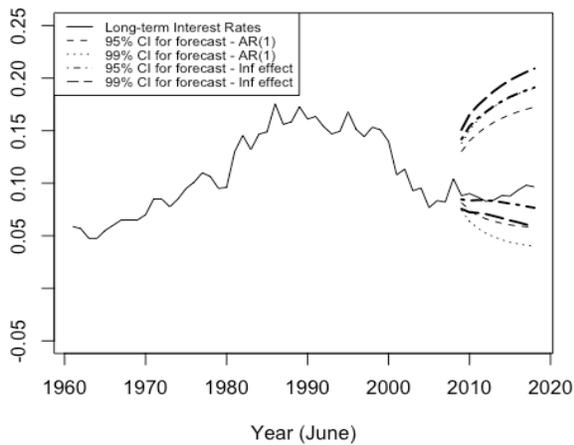
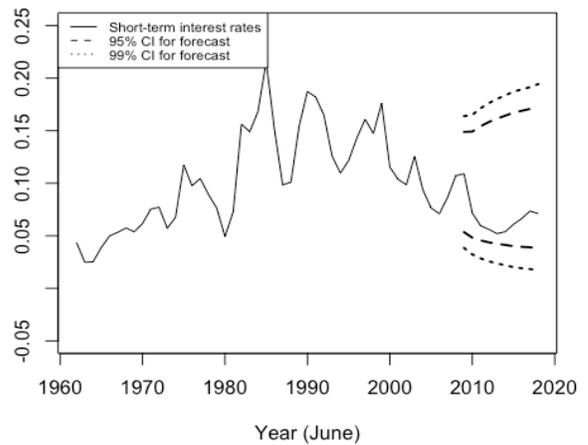

Figure 11. Backtesting Plots with 95% and 99% Forecast Intervals and Historical Data for Inflation, CPI, Dividend Yields, Dividends, Long-term and Short-term Interest Rates

## 10. PRACTICAL IMPLEMETATIONS

10.1 We believe that the model presented in this paper can easily be applied in practice. The intention is that simulations of key variables can be produced using simple software like Microsoft Excel. Whilst parameter estimates for all the variables are provided, we caution the practitioner against using all of these without adjustment.

10.2 For example, the model is calibrated to inflation data over the period 1959-2018. During this time there have been several distinct inflation regimes. It is therefore not advisable to simply use the mean parameter based on only historic data. In this instance, $\mu_q$ is 0.0809 when determined using historic data.

10.3 One can use a lower value for the long-term mean to at least be consistent with the inflation-targeting rather than the estimated mean parameter for forecasting purposes. The expected inflation parameter might be estimated from the appropriate prevailing bond curves. The approach is not prescriptive about the derivation process. However, it is important to keep the estimated value for the standard deviation to consider the uncertainty in the future inflation rates based on historical data.

10.4 Once the adjustment has been made in $\mu_q$, we suggest to run the simulations for all models and see how the chosen value affects the other models since price inflation play crucial role in the proposed cascade model. As can be inferred from backtesting simulations in the previous section, lower $\mu_q$ would produce lower forecast values for dividend yields and dividend growth as well as interest rates. The practitioners, then decide if any more adjustments are needed for long-term mean parameters for other economic series due to changing economic environment and/or policies.

## 11. CONCLUSIONS

11.1 In this paper we proposed a real world stochastic investment model for South African use. It is an updated and extended model proposed by Thomson (1996) which has a number of practical limitations due to lack of historical data. Our aim was to develop an economic scenario generator for long-term applications for both pension funds and life insurance companies.

11.2 We modelled different economic series by considering the bi-directional relations and also compared several models based on the statistical tests for independence and normality as well as the economic theory incorporated. We applied in-sample and out-of-sample model validation tests to analyse the performance of the proposed models. Except for the inflation-linked bonds model we examined the parameter stability of each economic series to comment on the robustness of the estimated values for long-term liability applications.

11.3 The price inflation has been modelled as an AR(1) process and influences the models for share dividends, dividend yields and long-term interest rates. The model is satisfactory based on both the statistical tests applied to the residuals and the forecast intervals calculated for backtesting. The dividend yields and share dividends model include moving average inflation effects while the data suggests that dividends also depend on previous year's dividend yields. The selected models satisfy in-sample model validation tests as well as producing reasonable forecasts compared to the historical observed values between 2008 and 2018. The long-term interest rates can be modelled as an AR(1) process although we also considered the moving average inflation effect. We examine the Fisher relation and rational expectation hypothesis considering the relation between the interest rates and moving average inflation effect derived from the model we proposed. The forecasts obtained from two models are quite close while the values produced by the model including the inflation effect are slightly higher. The short-term interest rates are modelled connected to the long-term interest rates and the model satisfies the necessary assumptions while funnel of doubt for the 10-year forecasts increases as the period extends. As for the inflation-linked bond rates, although several models fit the data satisfactorily as presented in the relevant section, the AR(1) model with an influence from short-term interest rates performs the best. Due to the short period of available data we could not analyse the forecasting aspects of the model for the inflation-linked bond rates.

11.4 As a further research we would like to discuss a practical application of the model and extend our model to incorporate real and nominal yield curves.

## 9. ACKNOWLEDGEMENTS

The authors deeply acknowledge the help of Prof. Robert J. Thomson and Prof. A. David Wilkie with regard to various aspects of this paper.

APPENDIX A – PARAMETER STABILITY

This Appendix presents the results for the parameter stability analysis of the dividend yields, share dividends, long-term and short-term interest rates models introduced in the paper.

Dividend Yields:

Figure 12 presents recursive estimates and 95% confidence intervals for the parameters $w_y$, $d_y$, $\mu_y$, $a_y$ and $\sigma_y$ for earlier sub-periods (data sets starting in 1985) and later sub-periods (data sets ending in 2018). Solid lines show the parameter estimates and the dotted lines show the 95% confidence intervals as explained in 3.6. Although we obtain relatively stable values for earlier and later sub-periods for $\mu_y$, $a_y$ and $\sigma_y$, the parameters which represent the inflation effect on dividend yields are not stationary particularly for shorter sub-periods. In 1980s and 1990s there are three jumps in $d_y$ parameter based on later sub-periods. We had to start with 25 years of data for the earlier and 22 years of data for the later sub-periods to have reasonable values due to the high number of parameters to be estimated.

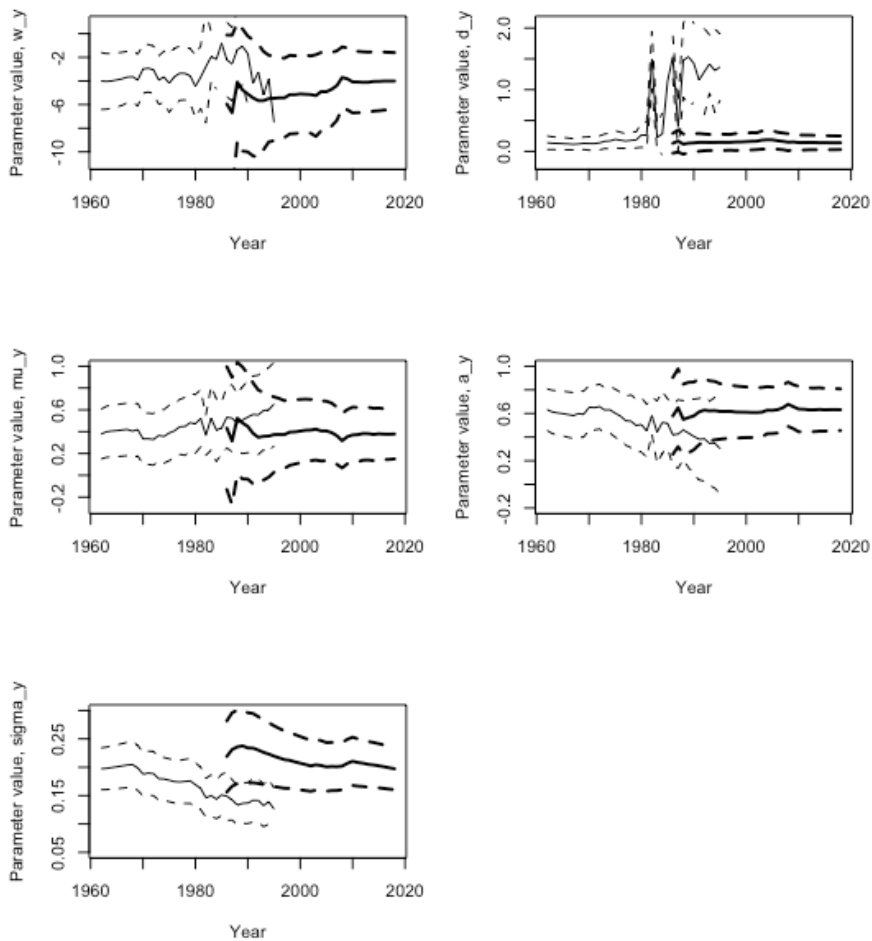

Figure 12. Estimates for parameters $w_y$, $d_y$, $\mu_y$, $a_y$ and $\sigma_y$ (solid lines show the parameter estimates and dotted lines show the 95% confidence intervals)

Share Dividends:

Share dividends model has six parameters to estimate since we need to insert the inflation and dividend yield effects. As the number of parameters increases more data is needed to obtain robust estimates. When we use shorter sub-periods to estimate all six parameters, it is not unusual to get high standard errors and hence wider confidence intervals. Figure 13 shows that the parameters representing the moving average inflation effect $w_d$ and $d_d$ have large standard errors. Overall mean $\mu_d$, standard deviation $\sigma_d$, dividend yield effect $y_d$ and moving average parameter $k_d$ have more stable estimates although there are a few jumps for some of the sub-periods.

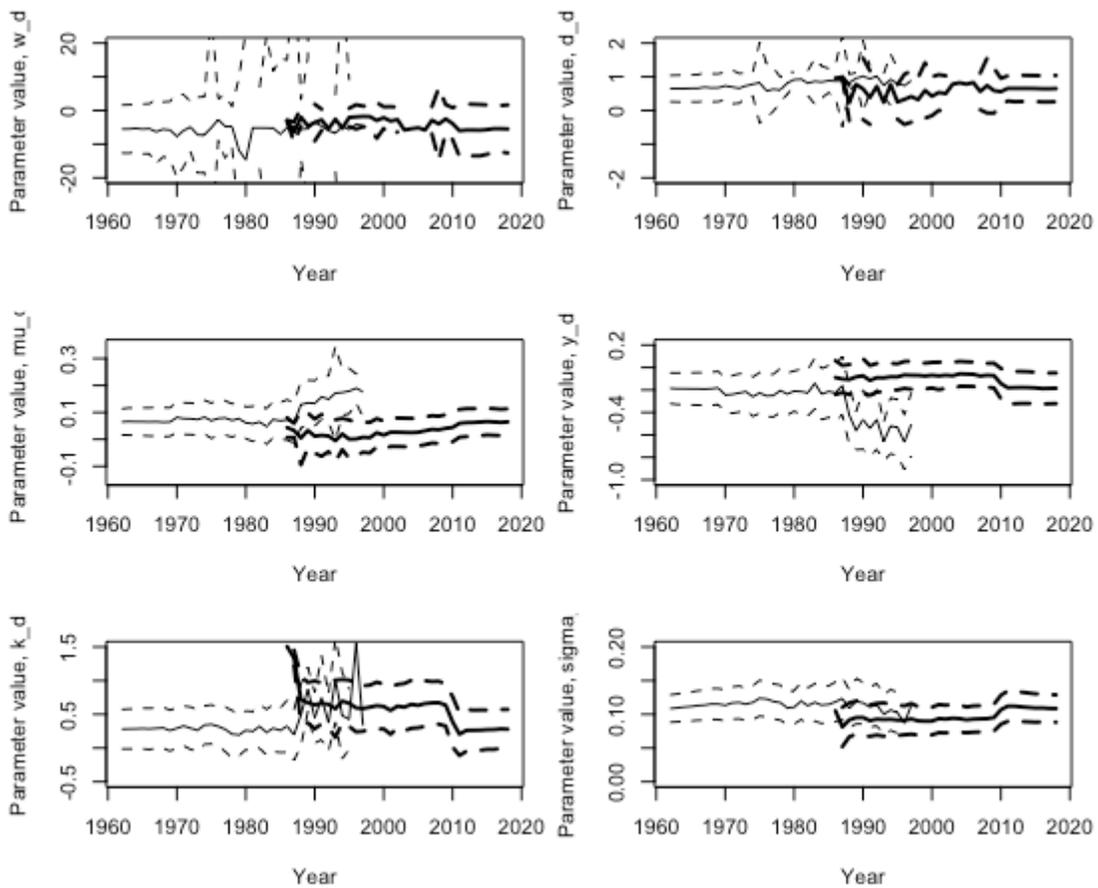

Figure 13. Estimates for parameters $w_d$, $d_d$, $\mu_d$, $y_d$, $k_d$ and $\sigma_d$ (solid lines show the parameter estimates and dotted lines show the 95% confidence intervals)

Long-term Interest Rates:

We examine the stability of the parameters of the long-term interest rates model by calculating the recursive estimates on incrementally larger data sets as we did in the previous sections. Figure 14 presents the recursive estimates and 95% confidence intervals of the mean level of long-term interest rates, $\mu_c$, the autoregressive parameter, $a_c$ and the standard deviation, $\sigma_c$ respectively, for the earlier (data sets starting in 1983) and later sub-periods (data sets ending in 2018). The values of most of the parameters are reasonably stable, except for $\mu_c$ which jumps around a lot for the later sub-periods, and $\sigma_c$, which has been increasing.

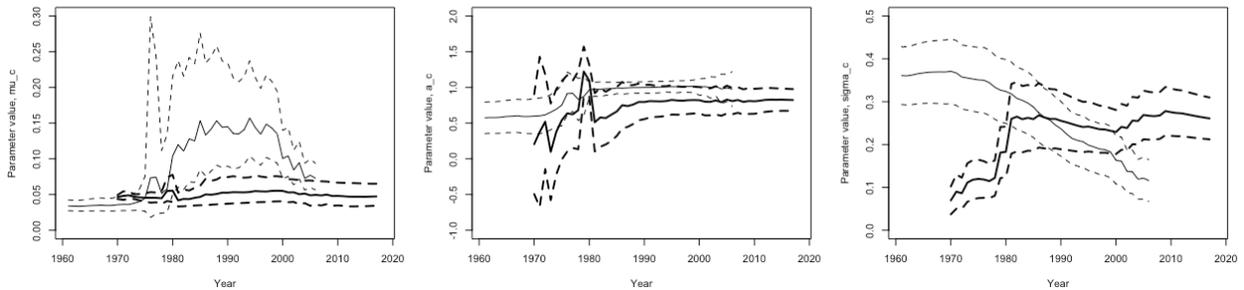

Figure 14. Estimates for parameters $\mu_c$, $a_c$ and $\sigma_c$ (solid lines show the parameter estimates and dotted lines show the 95% confidence intervals)

Short-term Interest Rates:

The stability of the parameters is examined using the same method as in previous sections. The values of $\mu_b$, $a_b$ and $\sigma_b$ over various sub-periods are shown in Figure 15. The graphs indicate that the parameters are quite stable over the whole period.

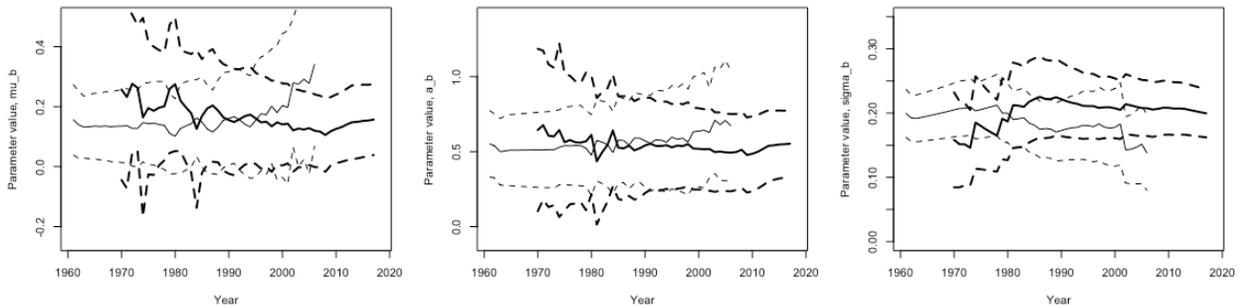

Figure 15. Estimates for parameters $\mu_b$, $a_b$ and $\sigma_b$ (solid lines show the parameter estimates and dotted lines show the 95% confidence intervals)

# APPENDIX B – BACKTESTING PARAMETERS

| Backtesting Parameters | Estimate | (Standard Error) |
|---|---|---|
| **Price Inflation (Fitted to 1960-2008)** | | |
| $\mu_q$ | 0.0951 | (0.0225) |
| $a_q$ | 0.8490 | (0.0711) |
| $\sigma_q$ | 0.0225 | (0.0023) |
| **Dividend Yields (Fitted to 1962-2008)** | | |
| $w_y$ | -3.7045 | (1.3039) |
| $d_y$ | 0.1482 | (0.0696) |
| $\mu_y$ | 0.3166 | (0.1259) |
| $a_y$ | 0.6780 | (0.0946) |
| $\sigma_y$ | 0.2020 | (0.0211) |
| **Share Dividends (Fitted to 1962-2008)** | | |
| $w_d$ | -2.8389 | (2.3418) |
| $d_d$ | 0.6082 | (0.3095) |
| $\mu_d$ | 0.0323 | (0.0249) |
| $y_d$ | -0.0650 | (0.0531) |
| $k_d$ | 0.6676 | (0.1602) |
| $\sigma_d$ | 0.0933 | (0.0097) |
| **Long-term Interest Rates (Fitted to 1961-2008)** | | |
| AR(1) Model | | |
| $\mu_c$ | -2.0809 | (0.0267) |
| $a_c$ | 0.9286 | (0.0460) |
| $\sigma_c$ | 0.0123 | (0.0012) |
| + MA Inflation Effect | | |
| $w_c$ | 1 | (fixed parameter) |
| $d_c$ | 0.13 | (fixed parameter) |
| $\ln\mu_c$ | -3.3452 | (0.1162) |
| $a_c$ | 0.5405 | (0.1238) |
| $\sigma_c$ | 0.3770 | (0.0389) |
| **Short-term Interest Rates (Fitted to 1961-2008)** | | |
| $\mu_b$ | 0.1055 | (0.0621) |
| $a_b$ | 0.5126 | (0.1281) |
| $\sigma_b$ | 0.2066 | (0.0213) |